\documentclass[preprint,1p,10pt,authoryear]{elsarticle}

\usepackage{color}

\usepackage{natbib}

\begin{document} 

\begin{frontmatter}
\title{Gender homophily from spatial behavior in a primary school: a sociometric study}

\author[cpt1,cpt2,ensae]{J. Stehl\'e\corref{cor1}}
\author[ensae]{F. Charbonnier}
\author[ensae]{T. Picard}
\author[isi]{C. Cattuto}
\author[cpt1,cpt2,isi]{A. Barrat}

\cortext[cor1]{Corresponding author}

\address[cpt1]{Aix Marseille Universit\'e, CNRS, CPT, UMR 7332, 13288 Marseille, France}
\address[cpt2]{Universit\'e de Toulon, CNRS, CPT, UMR 7332, 83957 La Garde, France}
\address[ensae]{ENSAE, 92245 Malakoff cedex, France}
\address[isi]{Data Science Laboratory, ISI Foundation, Torino, Italy}

\begin{abstract}  
We investigate gender homophily in the spatial proximity of children (6 to 12
years old) in a French primary school, using time-resolved data
on face-to-face proximity recorded by means of wearable sensors.
For strong ties, i.e.,
for pairs of children who interact more than a defined
threshold, we find statistical evidence of gender preference that increases with grade.
For weak ties, conversely, gender homophily is negatively correlated with grade
for girls, and positively correlated with grade for boys. This different evolution with grade
of weak and strong ties exposes a contrasted picture of gender homophily.
\end{abstract}


\end{frontmatter}

\noindent\textbf{Highlights}
\begin{itemize}
\item We investigate gender homophily in the spatial behavior of school children
\item We use behavioral data on face-to-face proximity measured by means of wearable sensors.
\item Strong ties provide evidence for gender homophily, which is slightly stronger for boys.
\item For strong ties, gender homophily increases with age.
\item For weak ties, gender homophily decreases with age for girls, while it increases for boys.
\end{itemize}
\hrulefill

\section*{Introduction}
The preference that individuals exhibit when they interact and build social ties with peers
they consider to be alike is a well known feature of human behavior
and is referred to as \textit{homophily} (see \cite{mcpherson} for a
review). The traits that may influence human relationships
are very diverse, ranging from physical attributes to tastes or political
opinions. The question of which similarities mostly shape social networks 
is rather open, and the answer seems to 
depend on age and on the nature of the considered social ties.
For example, gender or socio-ethnic homophily may vary over the lifespan of an individual,
and sharing working habits is more important among coworkers than among friends.
Together with other social mechanisms, such as triadic closure,
prestige or social balance~\citep{szell}, homophily shapes social relationships
and can play a role in the formation and stability of groups,
possibly influencing the constitution of social capital \citep{McDonald:2011}.
Nevertheless, it is often difficult to assess quantitatively to which
amount homophily contributes to social structures,
because many of the traits that are considered important are
changeable and might be modified by influence~\citep{Steglich}.
For example, similar smoking habits among teenagers may be important 
for the formation of new friendships, but at the same time
an individual with many friends who smoke might be influenced to become a smoker.
Unless panel data on social relationships and individual traits are available
it is impossible to disentangle the above effects~\citep{Kossinets:2009,Shalizi:2010},
with the exception of traits, such as gender, racio-ethnicity or social background, that are almost immutable.

Investigating the correlation between gender and interaction patterns,
therefore, allows to stay clear of the intricacies due to the interplay
of homophily and social influence.
A second advantage of studying homophily due to gender
lies in its universality: gender is among the most important traits that
shape interactions across cultures and, to different extents,
it plays a role during the entire life span~\citep{Mehta,Maccoby}.
Finally, gender homophily in social networks has been shown
to be linked to the broad issue of gender inequalities
on the job market~\citep{McDonald:2011}.
In the present analysis we will not make any distinction between sex and gender.

Here we study gender homophily among children as measured by monitoring their behavior in space.
Our work is based on high-resolution networks of spatial proximity
measured in a primary school in France over the course of two days.
The data were collected using wearable devices that sense and log
the face-to-face proximity relations of children over time.
The use of such unobtrusive devices yields detailed information
on the amount of time that children spend in close-range proximity,
allowing us to build a behavioral network representation
where nodes represent children and a link between two children indicates
that those children have spent time in face-to-face proximity (i.e., they have engaged in a ``contact'').
Each link can be weighted by the amount of time the corresponding individuals have spent in contact.
This network conveys information that is complementary
to the social ties assessed by means of surveys or questionnaires.
The time-varying proximity network captures objective information about one specific
behavior in space, automatically and without incurring in the recall biases of subjective reports
on spatial proximity to other individuals.
Surveys and questionnaires, on the other hand, capture a much broader class of ties,
and can integrate the full wealth of information on the subjective view of individual relationships.
Here we aim at building a fine-grained picture of gender-related behavioral homophily in children,
and leverage the detailed information on face-to-face proximity patterns provided by sensors
to produce a quantitative assessment of gender-dependent and age-dependence proximity behavior.
A natural question we want to address is the extent to which these behavioral findings
are congruent with the picture available in the literature, based on declared friendship relations
and on the observed actual behavior of children.

The article is structured as follows: a brief review of the literature
on gender homophily and its evolution through the lifespan is
presented first, together with some methodological perspectives on the
use of electronic devices to measure behavioral aspects of human
interactions. In the second part of the article we describe the
dataset and the methodology used to collect it.
Our results are presented in the third section. 
We report on statistical evidence of gender homophily in interaction patterns
(i.e., proximity patterns in space), and we discuss
the stability of neighborhoods from one day to the next.
Finally, we use the high temporal resolution of the data 
to investigate how the correlation of homophily with age depends on the (behaviorally-defined) strength
of the ties we consider.
We obtain contrasted results that shed new light on the evolution of weak ties with age.
We close with a short discussion and a call for further field research.

\section{Background}
\label{sec:background}
Quantitative analyses on gender preferences date back to Moreno's
seminal work on sociometry~\citep{WSS}, in which he introduces network
terminology to describe relations between children from kindergarten
through eighth grade. His study relies on direct observations and interviews with children,
and makes inferences about the variables that affect friendships.
He shows that although young children up to the second grade prefer same gender
mates, some of them also name friends of the opposite gender. 
This gender mixing then almost disappears, very few children making any mixed
friendship up to the sixth grade.
 
Several studies have since confirmed and extended these results.  A
recent review~\citep{Mehta} shows in particular a consensus about the
fact that gender homophily exists along the entire life span: it is
already present in infants' behaviors, increases up to a peak between 8
and 11 years~\citep{Maccoby}, in agreement with Moreno's study, and
decreases afterwards, mainly because of the development of romantic
relationships. It reaches a rather stable level among adults, although
studies on this life period remain scarce.
 
More recently, some differences of interaction styles between boys and
girls have been brought to light. The first and widely reported
difference is that boys tend to have a broader social network than
girls, who instead tend to make deeper and stronger relationships~\citep{Vigil,Lee}.
In particular, when children are asked to list their friends
with no limitation on the number, boys name more friends than girls
but most of the reciprocate nominations occur between girls. The
evolution of interaction styles is moreover different for both genders. 
\cite{LaFreniere} conclude from the 
direct observation of 193 children aged between 1 and 6 years 
that gender preferences increase earlier for girls than for boys,
but later on they become stronger for boys than for girls.
On the other hand, 
\cite{Martin} present a more moderate result.
In a direct observation of 61 children between 39 and 74 months of age, 
they do not observe a significant correlation between age
and the proportion of  same sex playmates for the entire sample,
but the correlation reaches a significant level when they consider boys only (this does not happen for girls). 
Indeed boys and girls behave differently, and differences in the amplitude of same gender
preference have also been observed. \cite{HT}
asked 186 children about their positive, neutral or negative attitude
toward all their classmates. While in the fourth grade
girls are more positive towards boys than boys towards girls, the
situation is reversed in the sixth grade.
\cite{Shrum} reach similar conclusions
from questionnaires identifying friendships.

During adolescence and the beginning of romantic relationships, girls
show an earlier evolution in their attitude toward other sex mates
than boys. \cite{Richards} report that girls declare to have
frequent thoughts about the opposite sex one grade earlier than boys
(but more moderately).  They also declare to spend nearly twice as
much time with boys as boys do with girls.  This
asymmetry may be explained by the fact that girls often have an older
boy as second best friend outside school, while boys rarely report
having girls as second best friends \citep{Poulin2007}.

Finally, stability of relationships, i.e. the maintenance of ties over
time, is a facet of friendship that has recently drawn some interest
(see \cite{Poulin2010} for a review). It is known to increase during
the primary school, which may be interpreted by the fact that concepts
such as reciprocity, loyalty and ability of solving difficulties
become more and more important in friendships at that age.
Relationships between same sex peers are also more stable than mixed
relationships but the empirical literature is still too shallow to
conclude about gender differences in terms of stability.
 
While most analysis in this field rely on questionnaires, surveys or
direct observations by adults, technological advances have led in the
recent years to the emergence of new tools to investigate human
behaviors and interaction patterns. In particular, wearable sensors
can now be used to detect
proximity~\citep{ONeill,Hui:2005,Pentland,Eagle,Salathe} and even
face-to-face spatial relations~\citep{Cattuto2010,SocioPatterns}. 

The study of the networks resulting from this kind of ties implies a
focus on \textit{behavioral} networks defined in terms of spatial
proximity, rather than on \textit{social} networks as defined from
friendship relations and built from questionnaires. 
These networks are known not to be completely disconnected:
in particular, it has been shown that children interact four times
more with members of their friendship group (identified with the
social-cognitive map procedure) than with
non-members~\citep{Gest}. Moreover, as noted by
\citet{Granovetter:1973}, social ties are not adequately described as
unidimensional relations, but are ``\textit{(probably a linear)
  combination of the amount of time, the emotional intensity, the
  intimacy (mutual confiding), and the reciprocal services which
  characterize the tie}''. The study of the relation between these
different ingredients of a social tie is however far from trivial, especially as
several of them are of qualitative nature.

In this respect, the use of wearable sensors embedded in unobtrusive wearable
badges~\citep{Cattuto2010} allows to focus on 
the amount of time spent together by individuals, and to largely
enhance our capabilities of quantitative studies of such
behavioral aspects of the relationships networks. On the one hand, the method is unsupervised and
does not require the continuous presence of observers, as in usual direct monitoring of behaviors. It has
in fact already been used to
monitor populations of several hundreds of individuals over several weeks~\citep{JTB}, a task which would have
been 
hardly possible with direct observation methods.
On the other hand,
it enables the precise, objective and reproducible definition of a tie
in terms of location and body posture: a tie at time $t$ between a
pair of individuals exists if they face each other in close proximity
\footnote{We emphasize that the wearable devices only record the
  face-to-face proximity of individuals, and not the possible
  occurrence of conversations or physical contacts: sustained
  face-to-face proximity is considered here as a behavioral proxy for
  the interaction between individuals.}. Note that such behavioral ties are by construction
reciprocal, while in the case of
friendship networks it is quite common that a declared friendship is
not reciprocated~\footnote{A non reciprocate tie obtained with a multiple
  name generator may be considered as a different perception of the
  relationship between the individuals, which can be interesting
  \textit{per se}, but it can also result from an informant bias (see
  \cite{Knoke} for a chapter on informant bias) leading to measurement
  errors.}.

The use of sensors allows to address quantitatively research questions
about homophily emerging from behavioral patterns,
in a way that is complementary to studies based on friendship networks.
We can investigate whether the friendship homophily is effectively translated
into behavioral homophily as measured by the amounts of time
children spend with same-gender and opposite-gender peers,
and whether behavioral homophily exhibits the same features as friendship homophily.
Some research questions can also be addressed in a more direct and objective way
than through surveys or diaries: for instance, as the measurement
can be carried out for several days, it is possible to measure directly
the similarity of behaviors from one day to the next,
and to compare the similarity of proximity patterns with same gender or opposite gender peers.
The direct observation of close-range proximity of individuals allows us to assign a well-defined
quantitative strength to each observed behavioral tie, for instance using the cumulated time that two individuals spend
in face-to-face proximity.
We can then define strong and weak ties
on the basis of a chosen threshold for this quantitative strength. The ensuing classification of observed ties into ``strong''
and ``weak'' ones is of course purely behavioral and it depends on an arbitrarily chosen threshold.
Regardless of the specific threshold value, such a classification achieves
an unambiguous discrimination of behavioral ties into interactions
that involve a strong commitment in terms of face-to-face presence
and interactions that, conversely, correspond to a short engagement in close-range proximity.
This makes it possible to contrast observations restricted to strong ties
with observations restricted to weak ties, exposing the differential role of such behaviorally-defined tie strength.
It also affords the investigation of specific properties of weak ties (as defined above),
a notoriously difficult task when relying on surveys, diaries or even direct observation.
Of course one can consider several other definitions of tie strength based on the
time-resolved proximity data we use. For example, the number of interactions between
individuals is another natural strength metric for ties, even though its 
relevance has been
debated in the context of several research domains \citep{Marsden2}.
The main argument is that contact frequencies are badly
correlated with the other measures of tie strength, mostly because of contextual
effects, which can be particularly relevant when considering interactions in a constrained environment~\citep{Webster}
\footnote{This issue is however probably rather limited in our study of the behavior of young children:
first, as explained below, we restrict the study to their 
behavior during the breaks and lunch time, where they can freely interact with anyone; second,
at the age considered, most of their friends are in fact schoolmates.}.
Here we focus on tie strength defined as the total (cumulated) time two individuals spend
in close-range proximity because it is one of the simplest tie strength definitions
we can define, it measures an intuitively important property of social behavior
(i.e., the amount of time spent in physical proximity),
and it hides the temporal complexity of multiple encounters,
which is known to be bursty and far from trivial to model~\citep{Gonzalez}.
Given this choice for the strength metric, we can study how the dependence of homophily with age
differs between strong ties (long face-to-face engagement) and weak ties (short face-to-face engagement) ties,
and how the time allocation to proximity in space differs across genders.

In the following, we investigate the above issues using high-resolution data on the
face-to-face proximity patterns of children in a primary school, measured over two days,
and originally collected in the context of an epidemiological study~\citep{PlosOne}.

\section{Data and methodology}
\label{sec:data}

Data on the face-to-face spatial proximity of more than $200$ school
children was collected during two consecutive days in October 2009 in
a primary school in Lyon, France (the school year starts in early
September). 
This school is located in a wealthy district of a large city and belongs to the private catholic sector. Participants were asked to wear proximity-sensing
electronic badges on their chest.  The badges engage in bidirectional
low-power radio communication, and can exchange radio packets only if
the individuals who wear them are facing each other within a
$1$m-$1.50$m range. Receiving stations (``readers'') located in the
school classes and in the playground collect real-time information
about the spatial proximity of participants. \cite{Cattuto2010}
provide detailed information on the sensing infrastructure,
and \cite{PlosOne} report a general analysis of the observed contact patterns.
The wearable sensors are tuned so that the face-to-face
proximity of two individuals wearing them can be assessed over an
interval of $20$ seconds with a probability in excess of $99$\%.
Two individuals are said to be in contact if their badges exchange
radio packets during a $20$-second time window,
and the contact is considered interrupted if the badges cannot exchange packets
over a $20$-second interval. Thus the minimum duration of a contact is $20$
seconds and all measured contact durations are multiples of $20$ seconds.
We emphasize once again that the sensors measure the face-to-face
proximity of individuals, and not the possible occurrence of
conversations or physical contacts.

The primary school under study is composed of $10$ classes, divided in
$5$ grades, labeled 1A, 1B, \dots, 5A, 5B (two classes for each grade).
The age of children ranges between 6 and 12 years.  The data were
collected over two school days, from 8:30am to 5:15pm. Only
interactions taking place on the school grounds were recorded. It is
worth noting that slightly more than one child in three leaves the
school premises for lunch, which leads to a relative drop of activity
during the lunch break. All of the $10$ teachers and $96$\% of the
children ($232$ out of $241$) took part in the data collection. The
$9$ remaining children were either missing on both days or received a
badge that was defective and had to be removed from the dataset.

\begin{table}
\begin{tabular}{ | c || c | c | c | c | }
\hline 
Class & Girls & Boys & Total & Class size\\
\hline 
1A & 11 & 10 & 21 & 24 \\
1B & 13 & 12 & 25 & 25 \\
2A & 14 & 9 & 23  & 25 \\
2B & 15 & 11 & 26 & 26 \\
3A & 9 & 14 & 23 & 24 \\
3B & 11 & 11 & 22 & 22 \\
4A & 8 & 11 & 19 &  23\\
4B & 10 & 13 & 23 & 24 \\
5A & 10 & 11 & 21 & 24\\
5B & 11 & 13 & 24 & 24 \\
\hline
\end{tabular}
\caption{Number of girls and boys in each class for which contact data and gender have been collected.}
\label{table:numbers}
\end{table}

Each individual is uniquely associated with one wearable badge and,
through that, to a unique numeric identifier. The identifier is only
associated with anonymous metadata for each individual: school class,
gender, year and month of birth. All the statistical treatment of the
data is performed in an anonymous way. The metadata was collected for
$227$ out of $232$ participating students. The difference is accounted
for by participants whose badge was accidentally replaced during the
deployment, breaking the connection between the badge identifier and
the participant metadata. We restrict our sample to this
subpopulation of $227$ children ($94$\% of the children in the
school), composed of $112$ girls and $115$ boys, and we do not consider the data relative to
teachers. We can reasonably assume the absence of any selection bias,
as the exclusion from the studied population is not related to gender
or behavior. Table~\ref{table:numbers} gives the corresponding numbers of boys and girls
in each class.

The proximity-sensing infrastructure records the time and duration of
each face-to-face proximity event (or ``contact'', in the following),
thus it is possible to construct an aggregated network of face-to-face
proximity over a given period of time -- for instance one day -- where
nodes represent individuals, and edges link individuals who have been
in contact at least once during the day. Each edge corresponds therefore to a sequence of contact events 
between the two linked individuals.

We do not have information on whether children conform to
pre-defined seating arrangements during class time, nor whether
teachers control or influence the seating patterns within classes. In
order to remove possible biases due to such factors, we study the
spatial proximity of children when they have maximum freedom of
associating with one another: we restrict our analysis to contacts
recorded in the playground and canteen of the school, which overall
account for $32\,027$ contact events ($41\%$ of the recorded
contacts). The recorded contacts are used to construct an aggregated
weighted network, in which nodes represent children and pairs of nodes
are connected by an edge if the corresponding pair of children has met
at least once during the two days. The weight $w_{ij}$ of an edge
between child $i$ and child $j$ is here defined as the cumulated duration
of all their contacts.  The resulting network comprises $227$ nodes
and $7070$ weighted edges.

\section{Results}
\label{sec:results}
\begin{figure}[ht]
	\begin{center}
    	\includegraphics*[width=10cm]{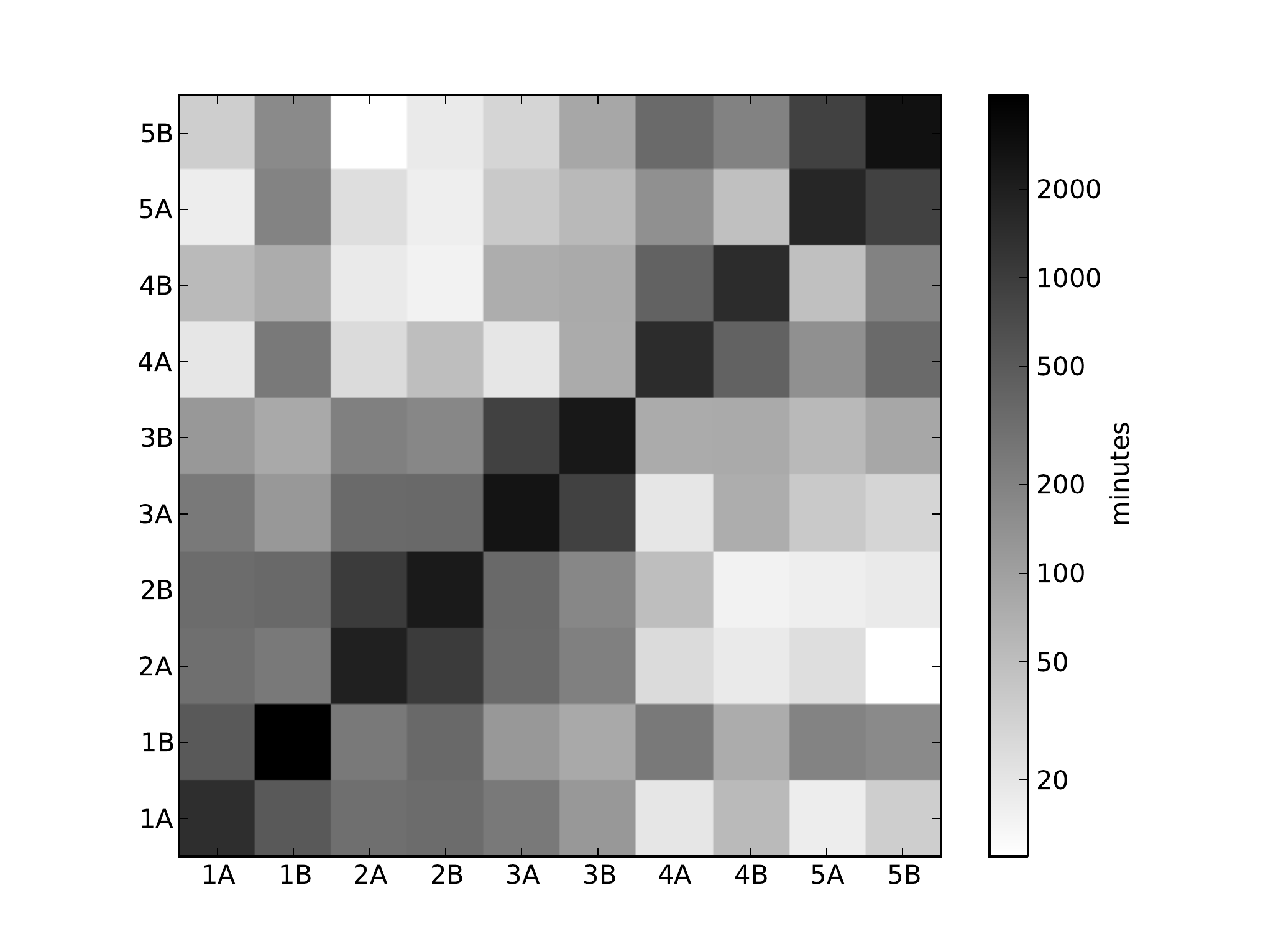}    
	\end{center}
  	\caption{Class-wise contact matrix: the matrix entry for row X
          and column Y indicates the total duration of all contacts
          recorded in the playground and in the canteen between class
          X and class Y, over the course of two days. The matrix
          values, expressed in minutes, are logarithmically
          greyscale-coded as shown in the legend.
	\label{fig:matrix}}
\end{figure}

Though we only consider contacts occurring in the playground or in the
canteen, the contact patterns are highly determined by the grade and
class structure of the school. Figure~\ref{fig:matrix} shows the
degree of contact between classes: each matrix element (row X, column Y)
is greyscale-coded according to the sum of contact durations
between all pairs of children belonging respectively to classes X and
Y, i.e., $W_{XY} = \sum_{i\in X} \sum_{j \in Y}w_{ij}$
(the weight $W_{XX}$ for contacts inside the same class X is twice the sum of all contact durations among the children of class X).
The high values along the diagonal of the matrix indicate that most contacts occur
between pairs of children belonging to the same class.
The division into grades is responsible for the visible block-diagonal structure
($2 \times 2$ darker squares along the diagonal).
Lower grades (1st to 3rd) also have less contact with higher grades (4th and 5th),
with the exception of one class in the 1st grade. Most of this structure
can be understood in terms of the schedule of class breaks and turns
at the canteen. In particular, in France, the first three grades (called ``cours pr\'eparatoire'' and
``cours \'el\'ementaire'') are rather clearly separated from the fourth and fifth grades (called ``cours moyen'').

\begin{figure}[ht]
	\begin{center}
	\includegraphics[width=0.7\textwidth]{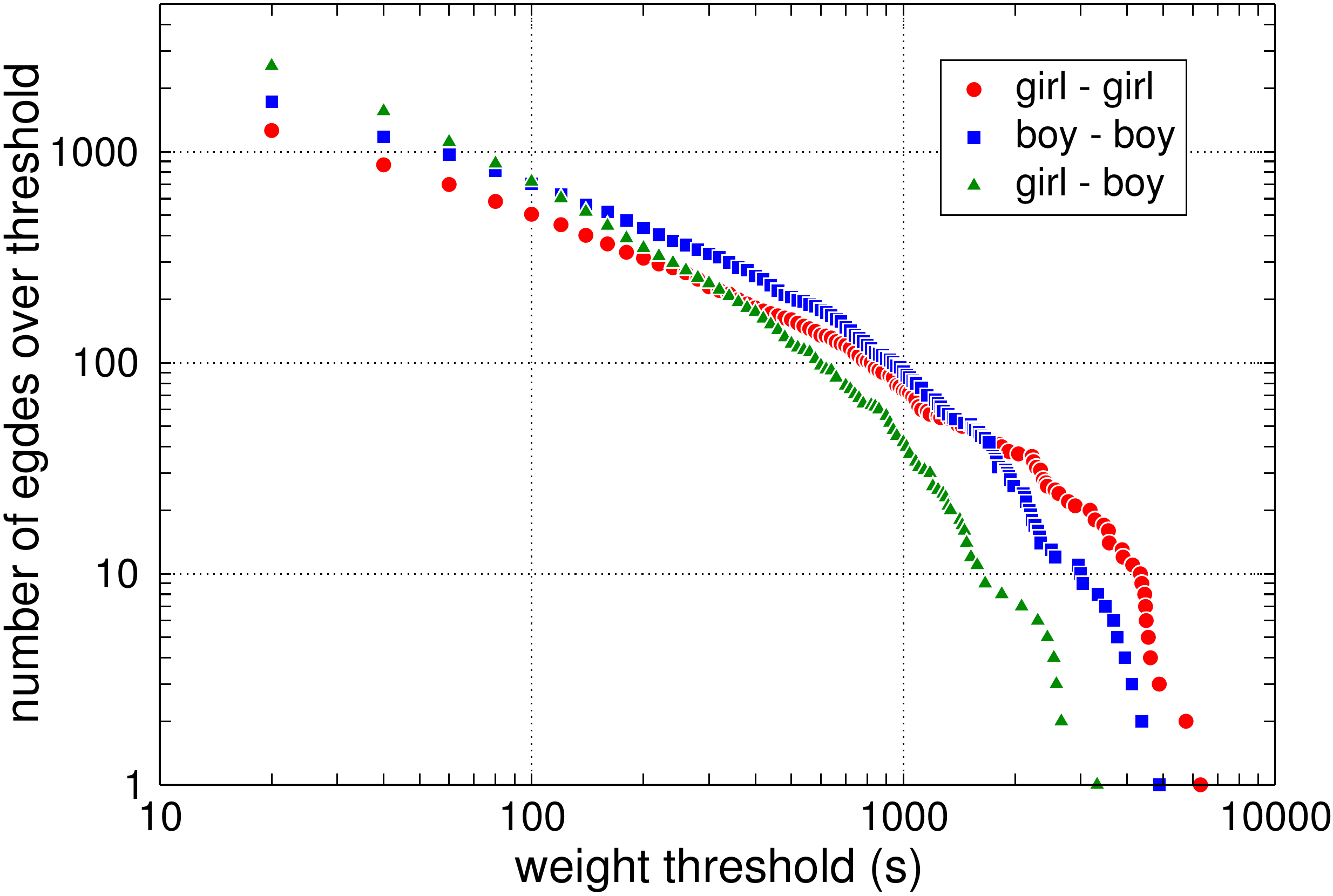}
	\caption{Cumulative histograms of edge weights of the contact
          network aggregated over two days. Edges are divided into
          three categories according to the gender of the connected
          nodes (boy--boy, boy--girl or girl--girl).
	\label{fig:weight_histo}}
	\end{center}
\end{figure}

Figure~\ref{fig:weight_histo} shows the cumulative weight histograms
for boy--boy, boy--girl or girl--girl edges in the aggregated contact network,
and provides a first indication of gender-related
differences. Edges between boys are more frequent than 
between girls ($1724$ vs $1261$) whilst there are almost as many girls
as boys ($N_g=112$ girls vs $N_b=115$ boys). There are more than
twice as many mixed edges than edges between girls ($2553$ vs $1261$),
but far less than twice the number of edges between boys.  The three
cumulative histograms shown in Fig.~\ref{fig:weight_histo} indicate
that mixed-gender edges tend to correspond to shorter cumulated
interactions relative to edges linking same-gender individuals: the
mean weights of mixed-gender and same-gender edges are $118$~s and
$242$~s, respectively.  Moreover, edges between girls have on average
higher weights than edges between boys, with mean weights of $265$~s
and $225$~s, respectively.

In this behavioral aggregated network the average degree is equal to
$62.3$, and has a higher value for boys than for girls ($67.1$ and
$57.3$, respectively). This difference is significant at the $10\%$
threshold (tested with a one-sided Wilcoxon test, $p=0.06$). When
considering the subgraph defined by edges with weight of at least 
$5$~min, the average degree is $8.0$, with a significantly higher
value for boys than for girls ($9.0$ vs $7.0$, $p=0.07$).

These two observations, that relate to the behavior of children,
are in agreement with results on friendship network structures,
and in particular with the literature about differences on group size and
level of intimacy, as reviewed by~\cite{Vigil}. They support the
hypothesis that men and women (here, boys and girls)
arbitrate differently between maintaining a large social group
and having more intimate and secure relationships.
Figure~\ref{fig:weight_histo} also clearly shows that cumulated contact durations are smoothly and broadly distributed,
extending over a large range of times. The behavioral data on face-to-face proximity lack
any intrinsically characteristic time scale, i.e., no typical duration can be defined for any type of contact.

\subsection{Statistical evidence of gender homophily}
In this section, we aim at testing statistically the evidence of
gender homophily. Our approach consists in comparing the
aggregated contact network with null models of graphs in which the probability
that an edge connects two nodes is independent of node genders, to
assess the probability that the observed contact network arises from such an
arrangement of contacts between individuals.

We first restrict the study to contacts occurring {\em within each class}:
the school schedule constrains contacts between classes, hence we
cannot assume that children have the same opportunity to make strong
ties within and across classes. Moreover, we only consider edges with
weight of at least $5$ minutes, i.e., whose contacts have a cumulated
duration of at least $5$ minutes over the two days of data. The
aggregated contact network has $531$ such edges, and in the following
we will indicate them as \emph{strong ties}, in reference to
Granovetter's terminology. The threshold of $5$ minutes is arbitrary:
it has been chosen to be large enough to eliminate \textit{weak ties}
that will be shown later (Sec.~\ref{sec:evolution}) to have different
properties with respect to gender homophily, and small enough to
retain a sufficient number of edges for statistical analysis. We have
checked that all of our results are robust with respect to changes in
the $5$ minutes threshold. We also note that the cumulated contact durations, the total
number of contact events, and the longest observed contact durations, are strongly
correlated, so that longer cumulated durations correspond  
to longer single contact events and that a filtering procedure on the cumulated duration 
or on the maximal duration of a contact would lead to the same results.

In order to test for homophily in a given class with $N_b$ boys and $N_g$ girls we consider the numbers of strong 
ties of each type: the total number $E$ of strong ties in the class is divided 
into $E_{gg}$ ties linking two girls, $E_{bb}$ ties linking two boys, and $E_{bg}$ ties involving a boy and a girl
($E = E_{gg}+E_{bb}+E_{bg}$).
Testing for homophilous behavior amounts to comparing the empirical
values of these parameters against the values they assume for null models in which the existence of a tie is not correlated with the gender of the individuals it links.

It is possible to design several null models of different complexity that control for different features of the empirical data. We begin by defining a null model that controls for the overall density of the contact network as well as for the difference in average degrees of boys and girls that was reported above.
We want the null model to have the same numbers $N_b$ and $N_g$ of boys and girls and the same number of edges 
$E$, i.e., the same density as the empirical contact network.
Moreover, let us denote the average degree of boys and girls 
respectively by $k_b = (2 E_{bb} + E_{bg})/N_b$ and $k_g = (2 E_{gg} + E_{bg})/N_g$. 
Since $N_b k_b + N_g k_g = 2E_{bb} + 2 E_{bg} + 2 E_{gg} = 2E$, a constraint on $k_b$ also determines $k_g$ (and vice versa). We can therefore
control for the different values of $k_b$ and $k_g$ by fixing, equivalently, $k_b$, $k_g$ or $k_b - k_g$. 
Here we choose the first solution.
Fixing $E$ and $k_b$ places two constraints on the three variables $E_{gg}$, $E_{bb}$ and $E_{bg}$, leaving the null model with a one dimensional random variable left, whose value will account for the degree of homophily in the allocation of the $E$ ties.

As it turns out, not only we can set $k_b$ (and hence $k_g$) to match the empirical values: we can actually design a null model that preserves the degree of each node, and thus, in particular, the average degrees $k_b$ and $k_g$.
To this aim, we construct 
an ensemble of random graphs with $N_b$ boys and $N_g$ girls having the same degrees as in the empirical
data by applying to the empirical network the reshuffling procedure
by \citet{Maslov:2004}\footnote{The
  procedure consists in taking random pairs of links $(i,j)$ and
  $(l,m)$ involving $4$ distinct nodes, and rewiring them as $(i,m)$
  and $(j,l)$. This is equivalent to the generation of a configuration
  model \citep{Molloy:1995} in which the degree sequence of the nodes is fixed to its
  empirical value.}, which preserves the degree of each node and hence
the average degree for every group of nodes.

In summary, our first null model corresponds to random graphs with
$N_b$ boys, $N_g$ girls, a fixed number of ties $E$, and a fixed
degree sequence (and as a consequence fixed average degrees $k_b$ and
$k_g$ for the boys and the girls).  We can therefore test for
homophilous behavior in the class by computing the distribution of the
fraction $E_{bg}/E$ of ties linking nodes of opposite gender in the
null model, and comparing it to the empirical value of this ratio.  To
this aim we consider the following null hypothesis:

{\it $H0$: the observed fraction $E_{bg}/E$ of ties involving a boy and a girl
  is compatible with that of a random
  graph in which nodes are labeled by gender ($N_b$ boys and $N_g$
  girls), the degree sequences is that of the empirical network,
  and the existence of a tie between two nodes does not depend on their gender.}

\begin{figure}[ht]
	\begin{center}
	\includegraphics[width=0.42\textwidth]{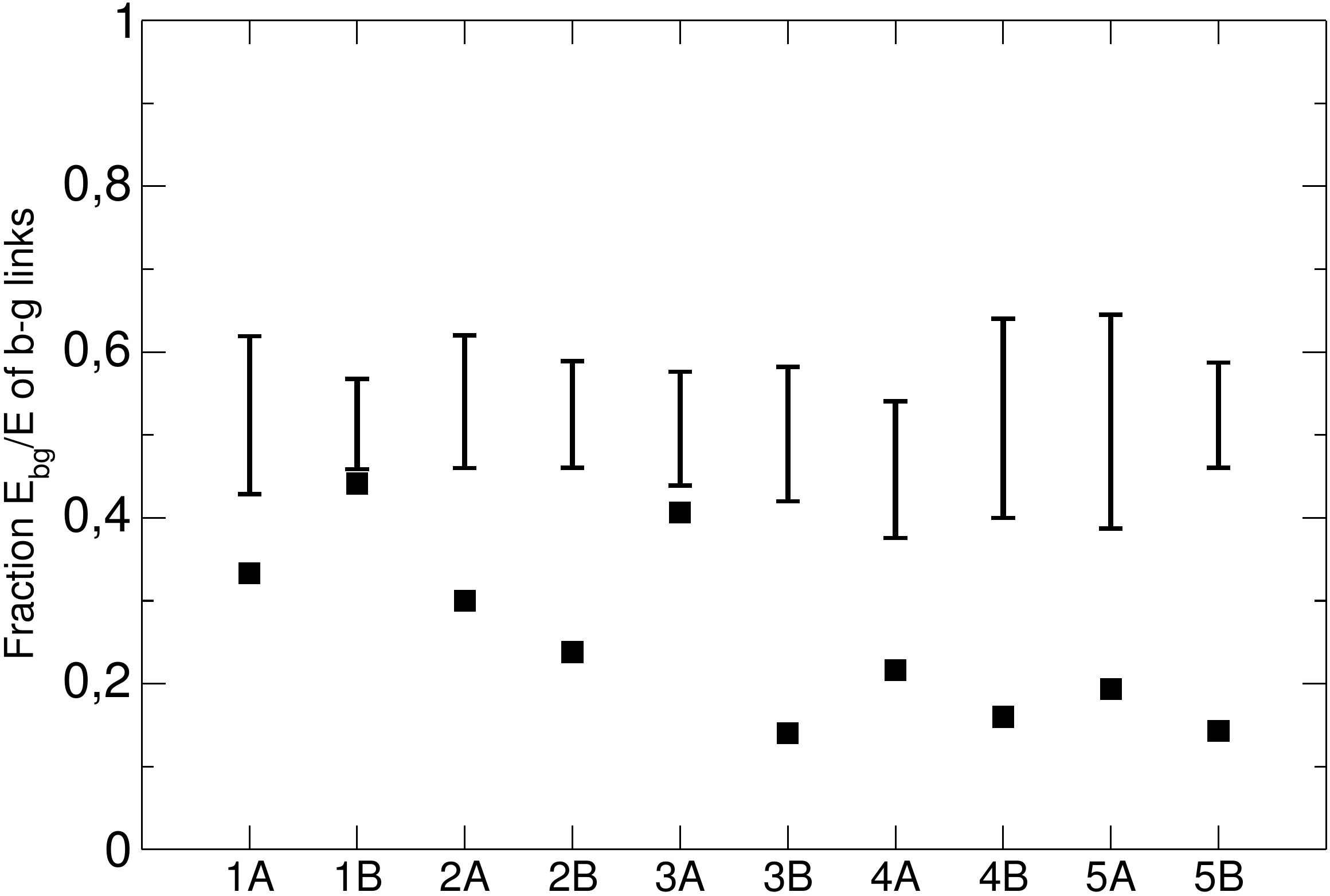}
        \hspace{0.5cm}
	\includegraphics[width=0.42\textwidth]{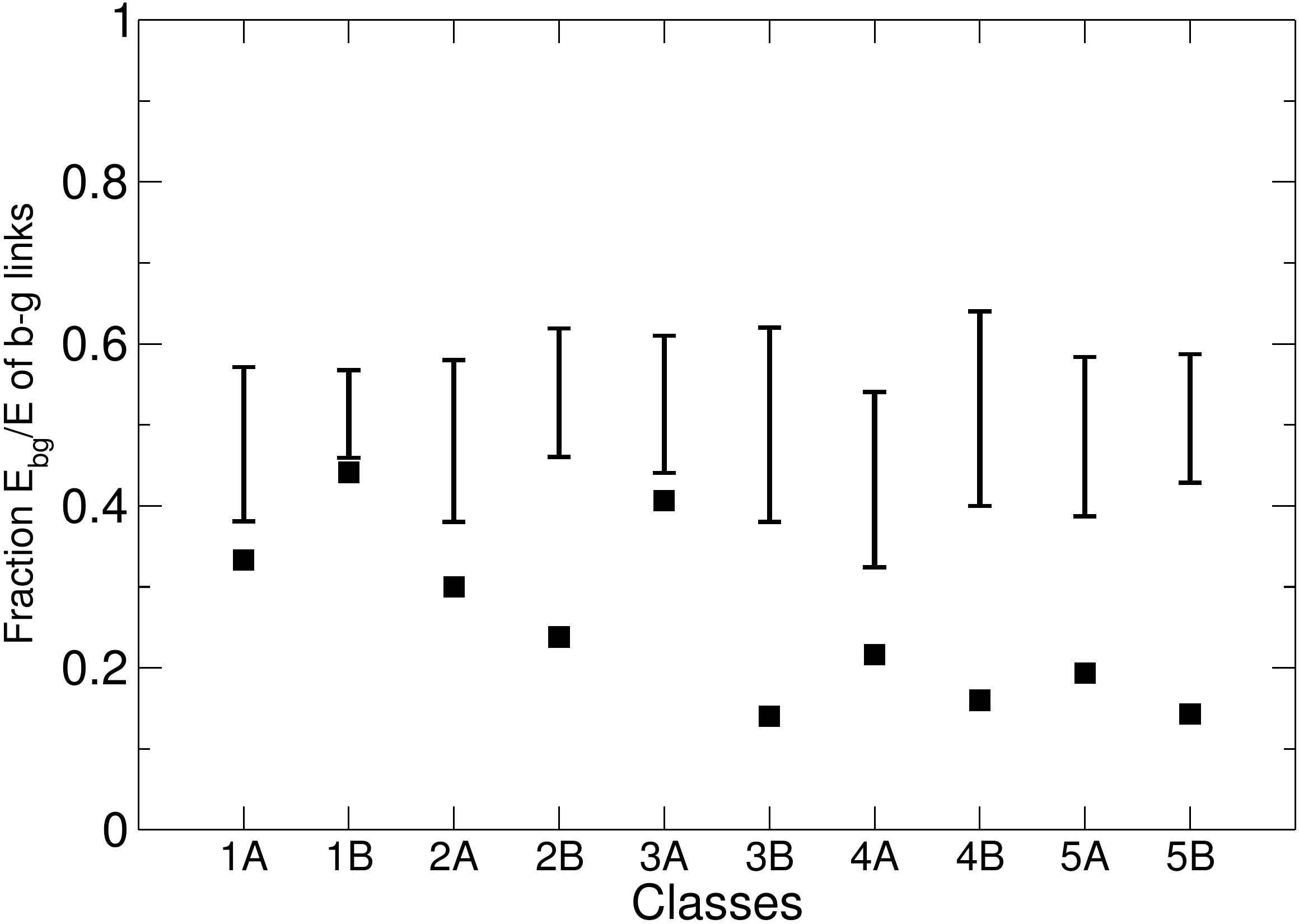}
	\caption{Statistical test of gender homophily for contacts
          within each class, restricted to contacts of cumulated
          duration of at least $5$ minutes over two days. Error bars
          indicate the $95\%$ confidence interval of acceptance of the
          null hypotheses of gender indifference. Symbols indicate the
          empirical fraction of ties involving a boy and a girl. Left:
          the error bars correspond, for each class,
          to the $95\%$ confidence interval
          for the values of $E_{bg}/E$ obtained under the null hypothesis $H0$.
          Right: same for the null hypothesis $H0'$.
	\label{fig:Ebg}}
	\end{center}
\end{figure}

The left-hand panel of
Fig.~\ref{fig:Ebg} shows for each class the region of acceptance
of the null hypothesis $H0$ at the $5\%$ threshold. The empirical
values are compatible with $H0$ in none of the classes, with values of
the fraction of ties joining students of different gender below the
$95\%$ confidence intervals. Hence we can reject the null hypothesis $H0$ of gender
independence of the edges.

It is however important to remark a limitation of the null model considered
above. The null hypothesis $H0$  disregards
all knowledge about the specific structure of the network
(the correlations between the numbers of neighbors of individuals in contact, the size of friendship clusters, etc.), except for the set of degrees of boys and girls.
In particular, 
a strong tie may exist between two
children independently of the fact that they may share many contacts.
It is on the contrary known that many triangles exist in the aggregated contact network, just like in many social networks.

To overcome this limitation, we design a different null hypothesis: we
consider the network of strong ties as fixed, and we randomly assign
the gender of each node, preserving the numbers of girls and boys in
each class. Under such a null hypothesis genders are interchangeable
and the network formed by the strong ties is independent from the
gender attributes. As previously discussed, we need to control for
the difference in the average degrees of boys and girls.
To this aim, we fix the network structure and we
consider all possible allocations of genders to nodes in which not
only the number of girls and boys are equal to their empirical values,
but also the average degree of boys (or equivalently of girls) is exactly equal to its
empirical value $k_b$. The corresponding null hypothesis can now be written as

{\it $H0'$: the observed fraction $E_{bg}/E$ of ties involving a boy and a girl is
  compatible with that of a graph having exactly the same topological
  structure as empirically observed, with $N_b$ nodes labeled as boys
  and $N_g$ as girls in a random fashion, in such a way that the average degree of
  boys is fixed to its empirical value (and hence the average degree of girls is also
  fixed to its empirical value).}

As shown in Fig.~\ref{fig:Ebg}, the results are very similar to the previous case and still provide statistical evidence for same-gender peer preference in all classes.

\subsection{Gender homophily from individual-based indices}

While the analysis of the previous paragraph was based on a global
measure (the number of same-gender ties inside a group), heterogeneity
between individuals can also be investigated through an individual
index. 

\subsubsection{Within class ties}

Let us first consider the same network as in the previous paragraphs,
i.e., restricted to strong ties (cumulated durations of contacts of at
least $5$ minutes) between children within each class.

\begin{figure}[ht]
	\begin{center}
	\includegraphics[width=0.7\textwidth]{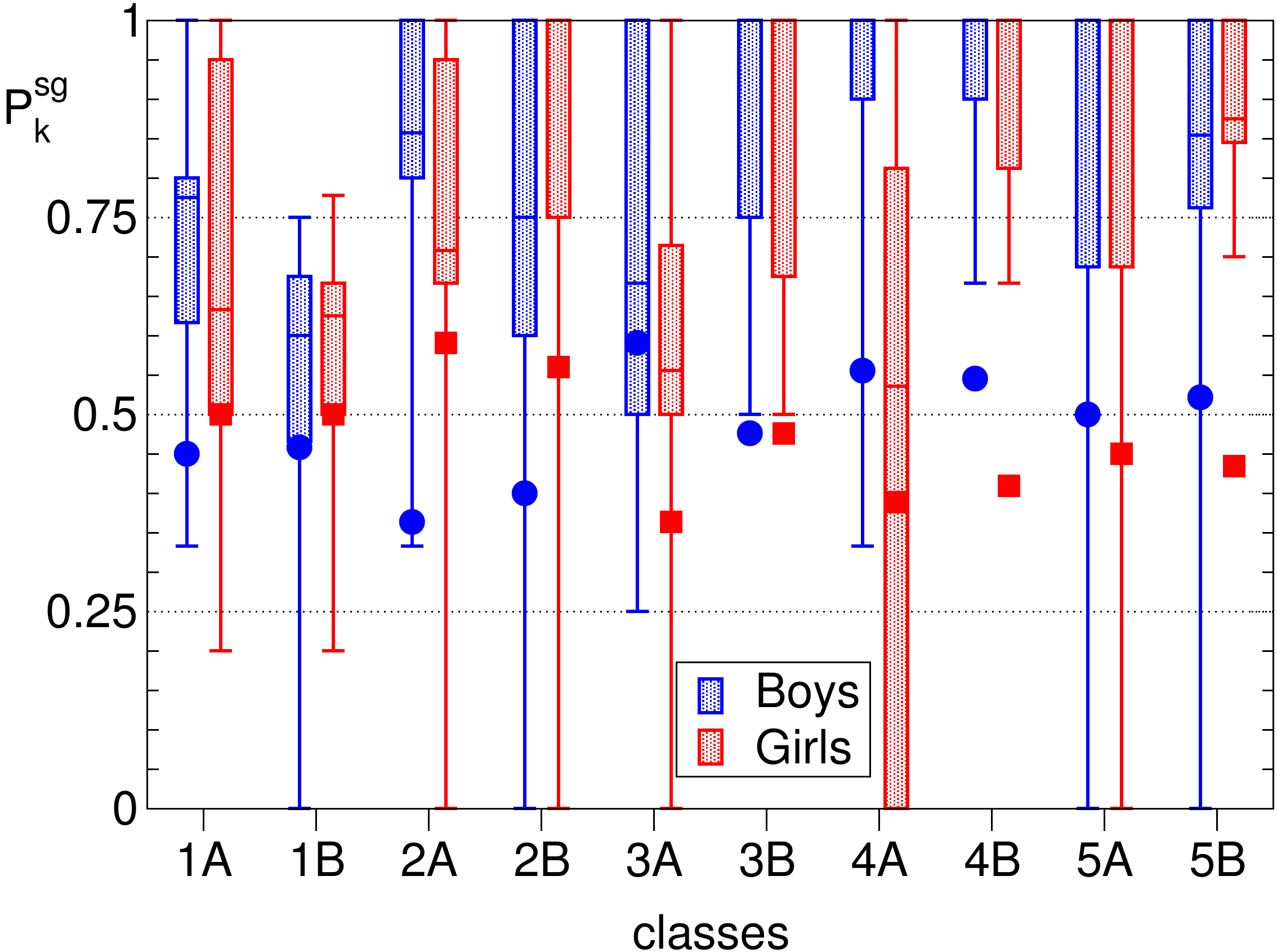}
	\caption{Boxplots of the same gender preference index
          $P_k^{sg}$, computed in each class separately for boys (blue) and girls
          (red). The center of the box indicates the median, its
          extremities the lower and upper quartiles, and whiskers
          indicate the smallest and largest values. The symbols
          indicate the baseline homophily for each gender in each
          class.
	\label{fig:pref_withinclass}}
	\end{center}
\end{figure}

For each node $i$ with at least one neighbor in this strong-tie
network ($215$ children out of the previous $227$), we compute the
proportion of same gender peers among its $k_i$ neighbors. We call
this index the individual homophily index, and we denote
it by $P_k^{sg}(i)$. This index is equal to $1$ if the considered
child has only same-sex neighbors and equal to $0$ if all the
neighbors are of the opposite sex.   
To interpret this index as a preference towards same-gender peers, 
the values of $P_k^{sg}(i)$ have to be compared with some baseline homophily. 
In the case of within class ties, this baseline homophily can be quantified with the ratios 
$(n_b -1) / (n_g+n_b-1)$ for
boys and $(n_g -1) / (n_g+n_b-1)$ for girls (the $-1$ is there because
an individual cannot have a link with him/herself), where $n_b$ and $n_g$ 
stand respectively for the number of boys and girls in the studied class.
These values take the gender ratio inside each class into account. 
Given any null model for which children would be indifferent to gender, 
the mean value of the individual homophily index would be given by these baselines.   
Figure \ref{fig:pref_withinclass} gives for each class the boxplot of the
individual homophily index, computed separately for boys
and for girls, together with the baseline homophily values.
We observe that most of the time, the baseline value is not  
contained between the first and the third quartiles of the empirical distribution of the index, 
thus illustrating the evidence of a gender homophily within classes as quantified by the distribution
of the individual homophily indices, even for small grades.  

\subsubsection{Inter- and intra-class ties}
\label{subsec:wholeschool}

We have considered so far ties within classes, corresponding to
interactions recorded in the playground or in the canteen. The class
structure makes it indeed easier to study
individual preferences of children to interact with same-sex peers: given that the 
children of a given class share the same schedule, we can 
assume that each child has an equal opportunity to interact for long (more than $5$
minutes) with same-class mates. This has allowed us to construct null
models and baseline homophily values to be compared with the empirical data.

However, the equal opportunity assumption has to be considered with
discernment. The setting allows to record only interactions within
the school premises.  It is very likely that some of these children
see each other during activities outside the school or in the
neighbourhood of their living places.  In that case, the opportunities
with these children would not be the same as with the others. If
the opportunities are correlated with gender (typically in the case of
sport activities), then the observed same-gender
preference would partly result of gender-correlated opportunities.
The main difficulty here comes from the fact that we do not ask
children to report on their preference but we deal with behavioral information on how they
interact with each other.  We now change our perspective in
considering only the proportion of same-gender mates these children
interact with during breaks and lunch, in a purely descriptive way,
relaxing the condition of being part of the same class.  We consider
the network composed of strong ties (cumulated durations of contacts
of at least $5$ minutes, as above) linking children belonging to
either the same class or different classes, for a total of $795$
ties~\footnote{In this case, children leaving the building for lunch
  will be less connected than others because they have a reduced
  opportunity to interact. We have checked that gender is independent
  from the behavior of eating at home, so that this does not bias our
  analysis.}.

\begin{figure}[ht]
	\begin{center}
	\includegraphics[width=0.7\textwidth]{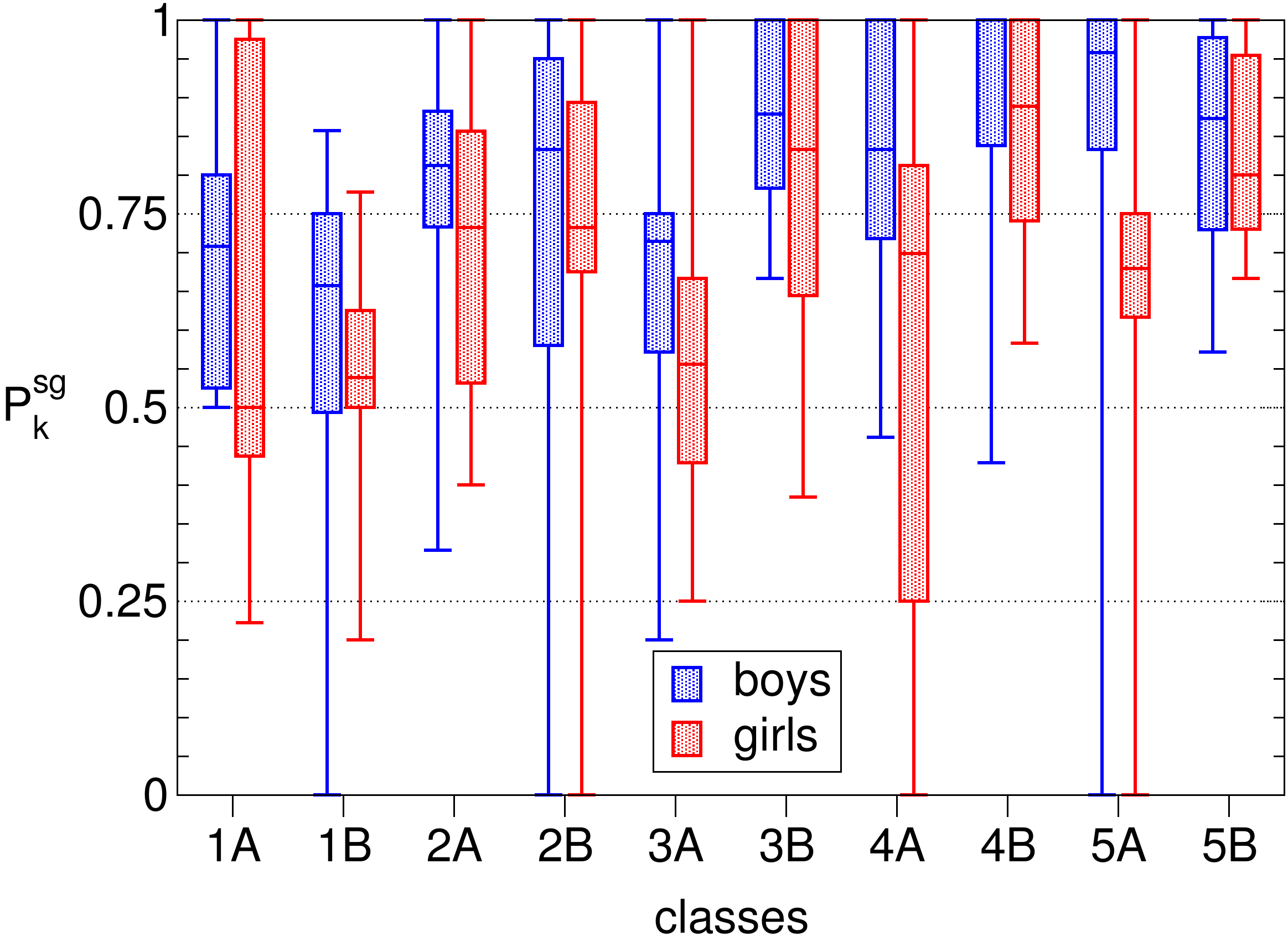}
	\caption{Same as Fig. \ref{fig:pref_withinclass} for the
          network of strong ties at the whole school level: Boxplots
          of the same gender preference index $P_k^{sg}$, computed
          separately for boys (blue) and girls (red). The center of
          the box indicates the median, its extremities the lower and
          upper quartiles, and whiskers indicate the smallest and
          largest values.
	\label{fig:pref}}
	\end{center}
\end{figure}

In this case, since the number of boys and girls in the school are
almost the same, the same gender preference index should be compared
to a gender-balanced neighborhood for which the index would take the
value $\approx 0.5$ ($112$ girls vs $115$ boys).  
Figure~\ref{fig:pref} reports the corresponding boxplots
for each class, separately for boys and for girls. The dispersion of
the index distribution is large, as indicated by the size of the box
and the whiskers.  While it happens that a child has no ties with
other children of the same gender, most values of the index are rather
high: same-gender homophily is present in all grades, for both
genders. Moreover, the figure seems to indicate that boys tend to
have a higher individual homophily index than girls, and that this
index increases with grades (we will examine this point in
Sec.~\ref{sec:evolution}).

The statistical difference between boys and girls can be estimated
through a one-sided Wilcoxon test. This non-parametric test is
preferred to a parametric one (such as ANOVA) because it does not
require any assumption on the form of the distribution of the
underlying random variable that generates the heterogeneity of
individual homophily index. We test the null hypothesis that the
averages of the individual homophily index are the same for boys and
girls, against the one-sided alternative hypothesis that the average
is higher for boys than for girls. For the 4th and the 5th grades the
null hypothesis is rejected at the $10\%$ threshold, meaning that
individual homophily is higher for boys and for girls
only for the two highest grades, but that, even in these cases, the effect
is only marginally significant.

\begin{figure}[ht]
\begin{center}
\includegraphics[width=0.7\textwidth]{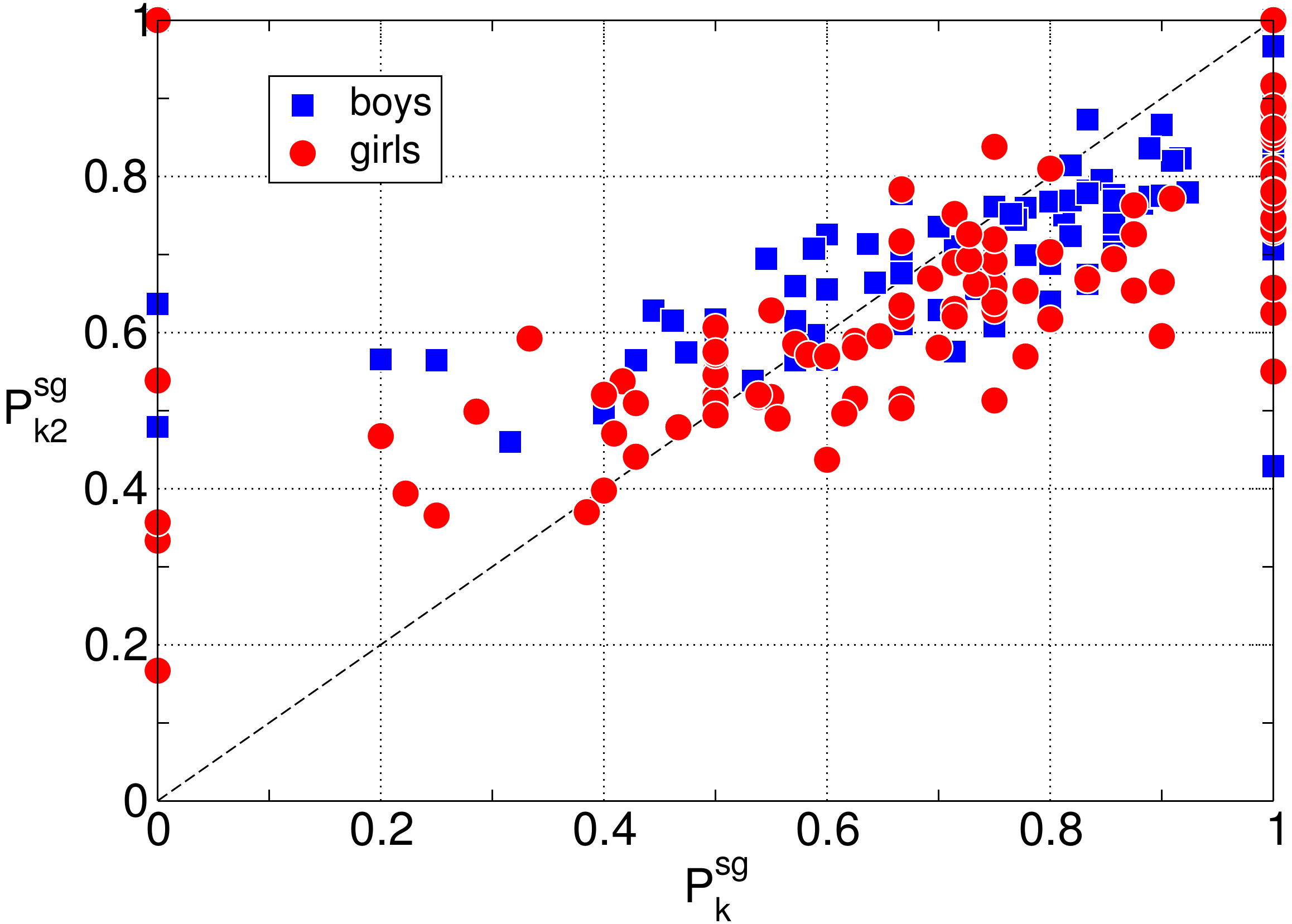}
\caption{Scatter plot of the two-step homophily index $P_{k2}^{sg}$ vs
  the individual homophily index $P_k^{sg}$, computed separately for
  boys (blue) and girls (red). The first bisector is represented by a
  dashed line.
\label{fig:pref_step2}}
\end{center}
\end{figure}

Interestingly, and despite the largest parts of the distributions
shown in Fig.~\ref{fig:pref} are above 1/2, some children have most of
their strong ties with children of the opposite sex.  The
interpretation proposed by Snijders~\footnote{Tom Snijders addressed
  this issue during the presentation he gave at the Universit\'e
  Paris-Dauphine when receiving his honorary doctorate on December 16,
  2011.}  is that this situation could be socially allowed under the
condition that the neighbors have as well many contacts with
individuals of the other sex.  To check this hypothesis, we define a
two-step homophily index for each node $i$, slightly modifying the
definition of the alters' covariate-average defined by \cite{RSIENA}:
$$
P_{k2}^{sg}(i) = \frac{1}{k_i} \sum_{j\in\mathcal{V}(i)}
\frac{k_j(\mbox{gender}_i)}{k_j}  \ .
$$  
In this expression, $\mathcal{V}(i)$ denotes the set of the $k_i$ neighbors of node $i$, 
$k_j$ is the number of neighbors of $j$, and $k_j(\mbox{gender}_i)$ is the number
of neighbors of $j$ having the same gender as $i$. $P_{k2}^{sg}(i)$ is therefore
the average
over the nodes $j$ belonging to the neighborhood $\mathcal{V}(i)$ of
node $i$, of the proportion of $j$'s neighbors who have the same
gender as node $i$.  Figure~\ref{fig:pref_step2} provides the scatter
plot of this two-step homophily index with respect to the previous
individual homophily index $P_{k}^{sg}(i)$.  Boys have on average a
higher two-step homophily index than girls (p-value $< 5 \cdot10^{-3}$
with a one-sided Wilcoxon test).  Moreover, if we consider egos having
a majority of neighbors of the opposite sex ($P_k^{sg} < 0.5$), the
neighbors themselves have on average more ties with children of the
same sex as ego than ego her/himself ($P_{k2}^{sg}(i) > P_k^{sg}$).

\subsection{Similarity of neighborhoods in different days}

We have focused in the previous paragraphs on an aggregated and
filtered view of the behavioral network of proximity between
children. In this view, we have considered the durations of contacts
aggregated over the whole measurement and used this information 
to define and focus on the strong ties. As the data is time-resolved,
it is however possible to compare the behavior of children in
different days, in a way that takes into account the contact durations
in a more complete way, and in order to understand the interplay
between gender homophily and the daily repetition of contact patterns.
We emphasize that we are here considering the similarity between 
the behavioral networks of children in successive days, rather than
the evolution (and possible decay) of social links in social networks,
that take place on longer timescales \citep{Burt:2000,Burt:2002}.

To this aim, we quantify the similarity between the neighborhood of each
individual~$i$ in day $1$ and day $2$ through the cosine similarity
\begin{equation}
\sigma (i) = \frac{\sum_j w_{ij,1} w_{ij,2}}{\sqrt{(\sum_j w_{ij,1}^2 ) (\sum_j w_{ij,2}^2) }} \, ,
\label{eq:sim}
\end{equation}
where the weight $w_{ij,1}$ is the cumulated time spent in face-to-face interaction
between $i$ and $j$ during day $1$, and $w_{ij,2}$ is the corresponding time during day $2$
(children absent in one of the two days are naturally excluded from this analysis).

We also consider two different similarity definitions that separately
measure the similarities of the same-gender and of the opposite-gender
neighborhoods across days ($\sigma_{sg}(i)$ and $\sigma_{og}(i)$,
respectively), by restricting the sums in Eq.~\ref{eq:sim}
to neighbors~$j$ who have the same (or the opposite) gender as $i$:
\begin{equation}
\sigma_{sg} (i) = \frac{\sum_{j \in {\cal V}_{sg}(i) } w_{ij,1} w_{ij,2}}{\sqrt{(\sum_{j \in {\cal V}_{sg}(i)} w_{ij,1}^2 ) (\sum_{j \in {\cal V}_{sg}(i)} w_{ij,2}^2) }} \, ,
\label{eq:sim_sg_b}
\end{equation}
where the sums over $j$ are restricted to the same-gender neighborhood ${\cal V}_{sg}(i)$ of $i$, 
and
\begin{equation}
\sigma_{og} (i) = \frac{\sum_{j \in {\cal V}_{og}(i) } w_{ij,1} w_{ij,2}}{\sqrt{(\sum_{j \in {\cal V}_{og}(i) } w_{ij,1}^2 ) (\sum_{j \in {\cal V}_{og}(i) } w_{ij,2}^2) }} \, ,
\label{eq:sim_og_b}
\end{equation}
where the sums over $j$ are restricted to the opposite-sex neighborhood ${\cal  V}_{og}(i)$ of $i$. 

\begin{figure}[ht]
\begin{center}
\includegraphics[width=0.7\textwidth]{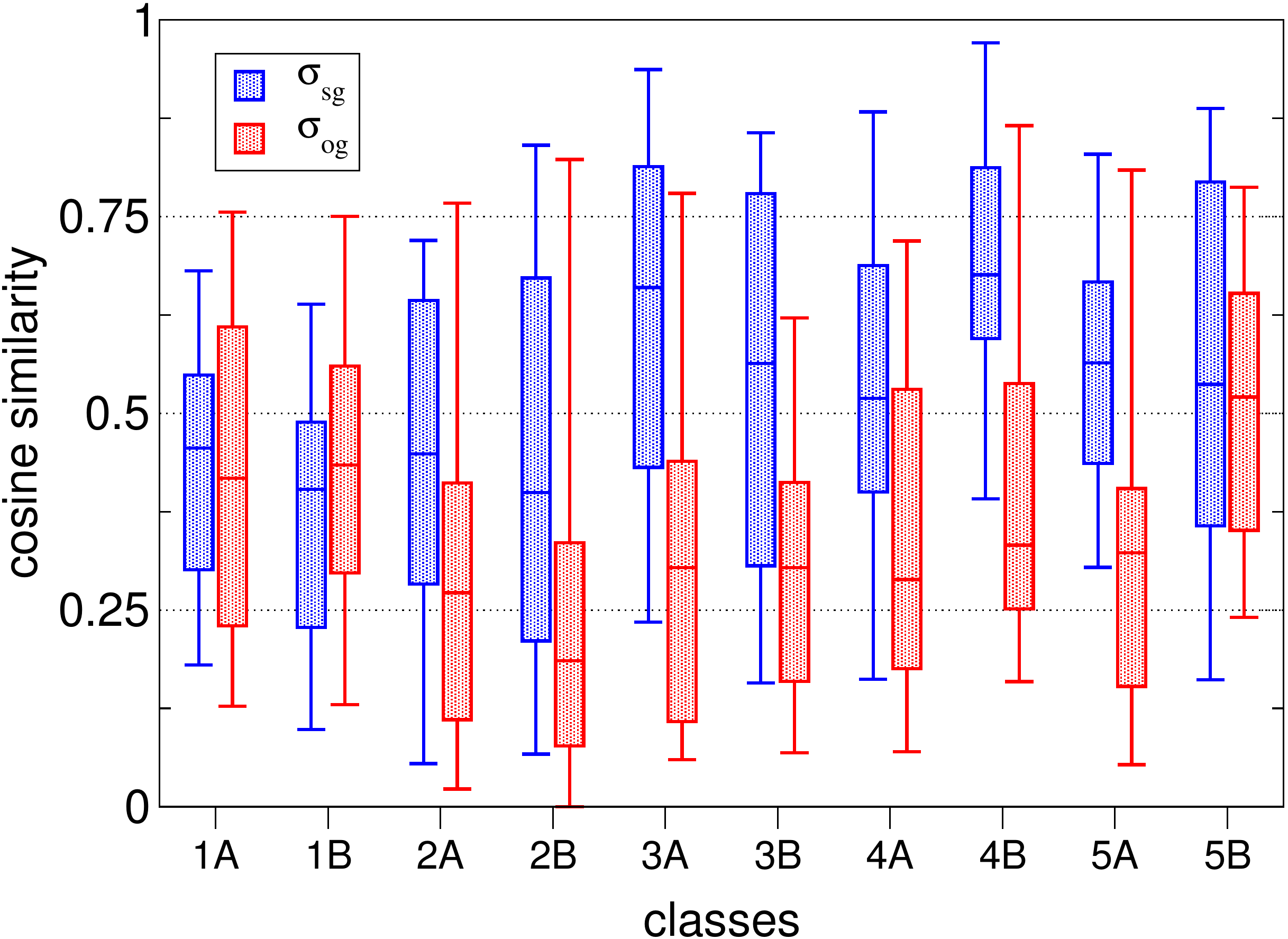}
\caption{Distributions of cosine similarities between the
  neighborhoods of the individuals of each class in days 1 and 2,
  restricted to same gender neighborhood $\sigma_{sg}$ and to opposite
  gender neighborhood $\sigma_{og}$.
\label{fig:cosine_sim}}
\end{center}
\end{figure}
Figure~\ref{fig:cosine_sim} displays the boxplots of the distributions
of $\sigma_{sg}$ and $\sigma_{og}$ for each class.  We test the null
hypothesis that the averages of $\sigma_{sg}$ and $\sigma_{og}$ are
the same against the one-sided alternative hypothesis that the average
is higher for $\sigma_{sg}$ than for $\sigma_{og}$.  Through a
Wilcoxon test, the null hypothesis is rejected at the $5\%$ threshold
for $6$ classes (one in the 2nd grade, one in the 5th grade, and all
classes of the 3rd and 4th grades), and with a p-value of $0.125$ for
the other class of the 2nd grade.  This shows that the same-gender
part of the neighborhood of an individual is statistically more stable
from day~1 to day~2 than the opposite-gender part of the neighborhood,
in agreement with the literature reviewed by \cite{Poulin2010}. On
the other hand, the stability of a child's neighborhood is not
significantly dependent on her/his gender: a Wilcoxon test of the null
hypothesis that the averages of $\sigma$ are different for boys and
girls leads to p-values larger than $0.3$ except for one class of the
third grade with $p=0.06$.

\subsection{Evolution of individual homophily with age}
\label{sec:evolution}

\begin{figure}[ht]
\begin{center}
\includegraphics[width=0.7\textwidth]{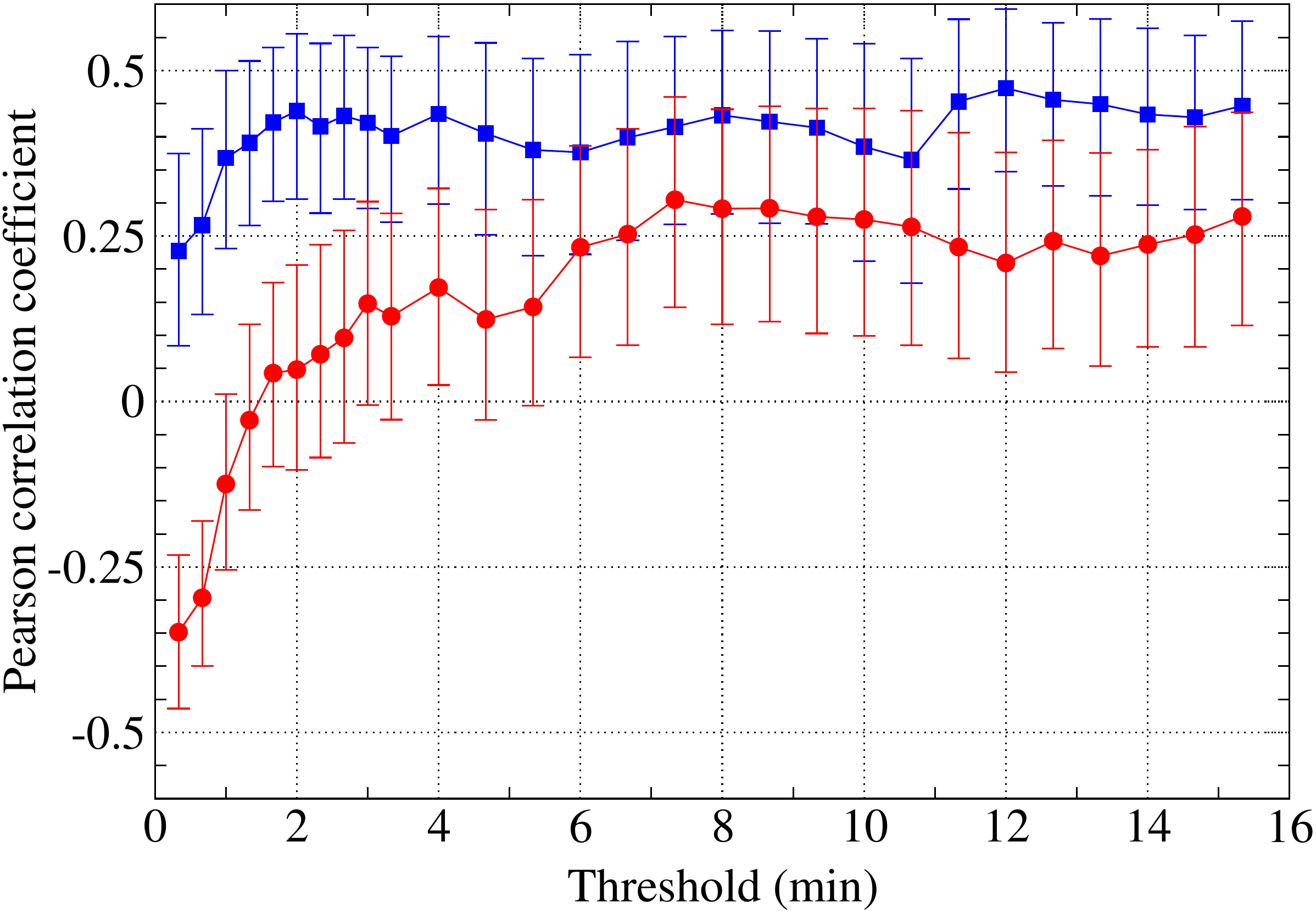}
\caption{Pearson correlation coefficients between age and same-gender
  homophily index $P_k^{sg}$, as a function of the threshold on edge
  weights. The correlation is computed separately for boys (blue
  squares) and for girls (red circles). Error bars indicate the
  bootstrap $90\%$ confidence interval, computed with $2000$
  resamples.
\label{fig:corr_by_age}}
\end{center}
\end{figure}

In subsection \ref{subsec:wholeschool}, we noted in
Fig.~\ref{fig:pref} a positive correlation between the same-preference
index and grade. This would mean that older children 
tend to have more contacts with
same-gender mates. This behavioral trend is in agreement with previous
studies focusing on social relations~\citep{Mehta,Maccoby,WSS}.
The information we have about the age of children allows us to explore this issue
in more detail, taking advantage of the quantitative temporal information encoded into the
behavioral network. We thus investigate
quantitatively the correlation between age and individual homophily
index, separately for boys and girls, and as a function of the weights of the links
considered. Figure~\ref{fig:corr_by_age}
shows the Pearson correlation coefficient between age and same-gender
preference index, as a function of the threshold on the cumulated
contact duration ($5$ minutes in the analysis above): when this
threshold is equal to $20$ seconds (the minimum duration of a contact)
we retain all ties, while on increasing it less and less edges are
kept (the strongest ones).

For both boys and girls, when we consider edges that correspond to a
cumulated interaction time of at least $100$ seconds (over $2$ days),
the correlation between the individual homophily index and age is
positive, and it is higher for boys than for girls.  However, when
weaker ties are retained (i.e., edges corresponding to shorter
cumulated times), the correlation is instead negative for girls.  This
means that the evolution of homophilous behavior with age is different
for weak and strong ties, and between boys and girls.  A closer
inspection reveals that for interactions of cumulated duration shorter
than $3$ minutes the number of same-gender mates decreases with age
for both genders, and that the number of opposite-gender neighbors
decreases even faster for boys, while it increases for girls.  For
contacts of cumulated duration larger than $5$ minutes, on the other
hand, the number of same-gender mates decreases for boys and stays
almost constant for girls, while the number of opposite-gender
neighbors decreases for both genders.  The increase in the number of
short encounters that girls have with boys may be related to their
earlier evolution in their attitude towards the other sex, as pointed
out by \cite{Richards} and \cite{Poulin2007}.  The same overall
picture emerges when analyzing the correlation of the same-gender
preference index with school grade rather than age, with a slightly
weaker correlation of individual homophily index and grade for boys.

As an aside, it may be noted that for high enough values of the
threshold some children become isolated in the network, meaning that
they have no interactions with other children that account for a
longer time than the threshold. This isolation does not affect equally
children who have either skipped or repeated a grade ($N=23$), and 
are thus younger (or older) than other children of the same class,
and the other children ($N=204$). At the $5$ minute threshold, $5$ children out of the
$23$ become isolated, compared to $7$ for those who have the same age
as their classmates. The relative risk for the children who have
either repeated or skipped a grade to become socially isolated,
compared to the other children, is equal to $6.33$ (a corresponding
odds ratio of $7.8$), indicating that children who skip or repeat grades 
might be more exposed to social exclusion.
However, the data collect took place relatively early in the year (Fall), at a time were 
some children are still new to others. It is thus plausible that
the social exclusion would be weaker later in the school year.

\section{Discussion}

The use of wearable sensors represents a new tool in the study and
description of child behavior. With respect to direct observation or
contact diaries, the presented methodology, based on unobstrusive
devices that can be worn during several consecutive days, has some
advantages. As it is unsupervised, it is much less limited in terms of
population size or duration of study and does not suffer from recall
biases. It also allows us to build a precise definition of a behavioral
tie, and gives access to quantitative information on the
duration of face-to-face proximity events between children. It yields
daily complete networks of face-to-face interactions of the school
population, with ties weighted according to behavioral information
(who spends time with whom), independently from recall biases.

It allows one therefore to investigate in some detail to which
extent features of friendship social networks, described in previous
literature mainly based on name-generator questionnaires, are
effectively recovered in terms of quantitative behavioral patterns.
Several features are indeed observed with statistical significance:
1) gender homophily is present in all grades of the primary school;
2) gender homophily reaches a higher level for boys than for girls in the 4th and 5th grades;
and 3) gender homophily tends to increase with age for strong ties,
at a higher rate for boys than for girls.

The quantification of face-to-face proximity behavior
allowed us to elucidate subtle aspects of behaviorally-defined gender homophily,
including gender-based and age-dependent properties.
In particular, we have shown that same-gender ties
are more similar across different days than mixed-gender ties,
which are subject to higher inter-day variability.
The unbiased discrimination of behaviorally-defined
strong and weak ties is a specific advantage of the behavioral proxy we selected:
information on weak ties in our framework is as accessible and reliable
as information about strong ties, in contrast to data obtained from surveys or diaries,
for which different kinds of cognitive and recall biases may apply depending on tie strength.
The definition of tie strength in terms of cumulated face-to-face time,
together with the objective empirical access to the time individuals spend in face-to-face proximity,
allowed us to probe how gender-based homophily in the proximity behavior
depends on tie strength.
This showed that, for strong ties, individual homophily
tends to increase with age at a higher rate for boys than for girls.
Conversely, for weak ties, the individual homophily is positively correlated with grade for boys,
while, remarkably, it is negatively correlated with grade for girls.

The empirical study of human interactions, and in particular the
influence of peer behaviors on individual outcomes is of major
interest in a broad range of social sciences, such as human behavior,
sociology, economy or education economy, and organizational
science~\citep{Manski}. The collection of empirical social network
data represents however a major barrier for the understanding of
social influences, especially in the context of models such as those
introduced by \cite{Doreian} or more recently by \cite{Steglich}. In this context, the
development of unsupervised methodologies that allow researchers to
collect large-scale, high-resolution dynamical data on human behavior
in a reproducible manner is a valuable asset that deserves some comments. 

As previously mentioned, new technologies entail important advantages in terms of ease of
deployment and measures at large scale and for consecutive
days. Re-test studies and comparisons can be easily carried out in
order to validate trends or to determine which features are
context-specific. It has of course to be underlined once again that
unsupervised detection of proximity patterns gives access to a
behavioral proxy rather than on real information about social
interactions (e.g., the different types of relationships are
disregarded). This behavioral proxy can be thought of as one of the
multiple components of a social relation, that can be here
quantitatively measured in a way that avoids informant biases such as
the limited recall of individuals about their contacts and
acquaintances. It represents an important tool to assess the similarity
between behavioral and friendship homophily, and, given this
similarity, represents a very interesting tool for further studies of
behavioral and social networks, in particular for the
study of the structure and evolution of weak ties.

The present study naturally stimulates further research
avenues. First, it would be interesting to investigate the
(quantitative) robustness of the results when different schools and
different moments of the school year are considered, and when a given
population is followed for longer periods. To this aim, deployments of
the infrastructure are planned for whole weeks, in different schools,
and in different periods of the year. The evolution of both age and
gender behavioral homophily along the school year could thus be
assessed. Moreover, more elaborate RFID badges endowed with an accelerometer
(under development) could allow us to test a possible
correlation between  gender homophily and different amounts of activity of boys and girls.
Finally, studies that would allow to make more accurate
comparisons between various ways of capturing social and behavioral
interactions would be of great interest. To this aim, detailed
comparison methods between automated and supervised data collection
methods, such as surveys, should be developed and are crucially
important to assess the specific limitations and potentials of the
methodology.

\newpage


\begin{thebibliography}{40}
\expandafter\ifx\csname natexlab\endcsname\relax\def\natexlab#1{#1}\fi
\expandafter\ifx\csname url\endcsname\relax
  \def\url#1{\texttt{#1}}\fi
\expandafter\ifx\csname urlprefix\endcsname\relax\def\urlprefix{URL }\fi

\bibitem[{Burt(2000)}]{Burt:2000}
Burt, R.~S., 2000. Decay functions. Social Networks 22~(1), 1 -- 28.

\bibitem[{Burt(2002)}]{Burt:2002}
Burt, R.~S., 2002. Bridge decay. Social Networks 24~(4), 333 -- 363.

\bibitem[{Cattuto et~al.(2010)Cattuto, Van~den Broeck, Barrat, Colizza, Pinton,
  and Vespignani}]{Cattuto2010}
Cattuto, C., Van~den Broeck, W., Barrat, A., Colizza, V., Pinton, J.-F.,
  Vespignani, A., Jul. 2010. Dynamics of person-to-person interactions from
  distributed rfid sensor networks. PLoS ONE 5~(7), e11596.
\newline\urlprefix\url{http://dx.doi.org/10.1371/journal.pone.0011596}

\bibitem[{Cynthia M.~Webster and Aufdemberg(2001)}]{Webster}
Cynthia M.~Webster, L. C.~F., Aufdemberg, C.~G., 2001. The impact of social
  context on interaction patterns. Journal of Social Structure.

\bibitem[{Doreian(1980)}]{Doreian}
Doreian, P., 1980. Linear models with spatially distributed data. Sociological
  Methods \& Research 9~(1), 29--60.

\bibitem[{Eagle et~al.(2009)Eagle, Pentland, and Lazer}]{Eagle}
Eagle, N., Pentland, A., Lazer, D., 2009. Inferring social network structure
  using mobile phone data. Proceedings of National Academy of Sciences
  106~(36), 15274--15278.

\bibitem[{Gest et~al.(2003)Gest, Farmer, Cairns, and Xie}]{Gest}
Gest, S.~D., Farmer, T.~W., Cairns, B.~D., Xie, H., 2003. Identifying
  children's peer social networks in school classrooms: Links between peer
  reports and observed interactions. Social Development 12~(4), 513--529.

\bibitem[{Gonz\'alez et~al.(2008)Gonz\'alez, Hidalgo, and Barabasi}]{Gonzalez}
Gonz\'alez, M.~C., Hidalgo, C.~A., Barabasi, A.~L., 2008. Understanding
  individual human mobility patterns. Nature 453, 479.
\newline\urlprefix\url{doi:10.1038/nature06958}

\bibitem[{Granovetter(1973)}]{Granovetter:1973}
Granovetter, M., 1973. The strength of weak ties. American Journal of Sociology
  78, 1360 -- 1380.

\bibitem[{Hayden-Thomson et~al.(1987)Hayden-Thomson, Rubin, and Hymel}]{HT}
Hayden-Thomson, L., Rubin, K.~H., Hymel, S., 1987. Sex preferences in
  sociometric choices. Developmental Psychology 23~(4), 558 --562.

\bibitem[{Hui et~al.(2005)Hui, Chaintreau, Scott, Gass, Crowcroft, and
  Diot}]{Hui:2005}
Hui, P., Chaintreau, A., Scott, J., Gass, R., Crowcroft, J., Diot, C., 2005.
  Pocket switched networks and human mobility in conference environments. In:
  WDTN '05: Proceedings of the 2005 ACM SIGCOMM workshop on Delay-tolerant
  networking. ACM, New York, NY, USA, pp. 244--251.

\bibitem[{Isella et~al.(2010)Isella, Stehl\'e, Barrat, Cattuto, Pinton, and
  Broeck}]{JTB}
Isella, L., Stehl\'e, J., Barrat, A., Cattuto, C., Pinton, J.-F., Broeck, W.
  V.~D., 2010. Whatʼs in a crowd? analysis of face-to-face behavioral
  networks. Journal of Theoretical Biology 271, 166--180.

\bibitem[{Knoke and Yang(2008)}]{Knoke}
Knoke, D., Yang, S., 2008. Social network analysis, 2nd Edition. No. 07-154 in
  Quantitative applications in the social sciences. Sage Publications.

\bibitem[{Kossinets and Watts(2009)}]{Kossinets:2009}
Kossinets, G., Watts, D.~J., 2009. Origins of homophily in an evolving social
  network. American Journal of Sociology 115, 405--450.

\bibitem[{La~Freniere et~al.(1984)La~Freniere, Strayer, and
  Gauthier}]{LaFreniere}
La~Freniere, P., Strayer, F.~F., Gauthier, R., 1984. The emergence of same-sex
  affiliative preference among preschool peers: a developmental/ethological
  perspective. Child development 55~(5), 1958--1965.

\bibitem[{Lee et~al.(2007)Lee, Howes, and Chamberlain}]{Lee}
Lee, L., Howes, C., Chamberlain, B., July 2007. Ethnic heterogeneity of social
  networks and cross-ethnic friendships of elementary school boys and girls.
  Merril-Palmer Quaterly 53~(3), 325 -- 346.

\bibitem[{Maccoby(2003)}]{Maccoby}
Maccoby, E.~E., 2003. The Two Sexes: Growing Up Apart, Coming Together, 5th
  Edition. Belknap Press of Harvard University Press.

\bibitem[{Manski(1993)}]{Manski}
Manski, C.~F., 1993. Identification of endogenous social effects: The
  reflection problem. The Review of Economic Studies 60~(3), 531--542.

\bibitem[{Marsden and Campbell(2012)}]{Marsden2}
Marsden, P.~V., Campbell, K.~E., 2012. Reflections on conceptualizing and
  measuring tie strength. Social Forces 91~(1), 17--23.
\newline\urlprefix\url{http://sf.oxfordjournals.org/content/91/1/17.short}

\bibitem[{Martin and Fabes(2001)}]{Martin}
Martin, C.~L., Fabes, R.~A., 2001. The stability and consequences of young
  children's same-sex peer interactions. Developmental Psychology 37~(3), 431
  -- 446.

\bibitem[{Maslov et~al.(2004)Maslov, Sneppen, and Zaliznyak}]{Maslov:2004}
Maslov, S., Sneppen, K., Zaliznyak, A., 2004. Detection of topological patterns
  in complex networks: correlation profile of the {I}nternet. Physica A 333,
  529--540.

\bibitem[{McDonald(2011)}]{McDonald:2011}
McDonald, S., 2011. What's in the “old boys” network? accessing social
  capital in gendered and racialized networks. Social Networks 33~(4), 317 --
  330.

\bibitem[{McPherson et~al.(2001)McPherson, Smith-Lovin, and Cook}]{mcpherson}
McPherson, M., Smith-Lovin, L., Cook, J.~M., 2001. Birds of a feather:
  Homophily in social networks. Annual Review of Sociology 27, 415--445.

\bibitem[{Mehta and Strough(2009)}]{Mehta}
Mehta, C.~M., Strough, J., 2009. Sex segregation in friendships and normative
  contexts across the life span. Developmental Review 29~(3), 201 -- 220.

\bibitem[{Molloy and Reed(1995)}]{Molloy:1995}
Molloy, M., Reed, B., 1995. A critical point for random graphs with a given
  degree sequence. Random Struct. Algorithms 6, 161.

\bibitem[{Moreno(1953)}]{WSS}
Moreno, J.~L., 1953. Who shall survive? Foundations of sociometry, group
  psychotherapy and socio-drama, 2nd Edition. Oxford, England: Beacon House.

\bibitem[{O'Neill et~al.(2006)O'Neill, Kostakos, Kindberg, gen. Schieck, Penn,
  Fraser, and Jones}]{ONeill}
O'Neill, E., Kostakos, V., Kindberg, T., gen. Schieck, A.~F., Penn, A., Fraser,
  D.~S., Jones, T., 2006. Instrumenting the city: Developing methods for
  observing and understanding the digital cityscape. Lecture Notes in Computer
  Science 4206, 315.

\bibitem[{Pentland(2007)}]{Pentland}
Pentland, A., May 2007. Automatic mapping and modeling of human networks.
  Physica A: Statistical Mechanics and its Applications 378~(1), 59--67.

\bibitem[{Poulin and Chan(2010)}]{Poulin2010}
Poulin, F., Chan, A., 2010. Friendship stability and change in childhood and
  adolescence. Developmental Review 30, 257 -- 272.

\bibitem[{Poulin and Pedersen(2007)}]{Poulin2007}
Poulin, F., Pedersen, S., 2007. Developmental changes in gender composition of
  friendship networks in adolescent girls and boys. Developmental Psychology
  43~(6), 1484--1496.

\bibitem[{Richards et~al.(1998)Richards, Crowe, Larson, and Swarr}]{Richards}
Richards, M.~H., Crowe, P.~A., Larson, R., Swarr, A., 1998. Developmental
  patterns and gender differences in the experience of peer companionship
  during adolescence. Child Development 69~(1), 154--163.

\bibitem[{Ripley et~al.(2011)Ripley, Snijders, and Preciado}]{RSIENA}
Ripley, R.~M., Snijders, T.~A., Preciado, P., 2011. Manual for siena version
  4.0. Oxford: University of Oxford, Department of Statistics; Nuffield
  College, (version January 17, 2012).

\bibitem[{{Salathe} et~al.(2010){Salathe}, {Kazandjieva}, {Lee}, {Levis},
  {Feldman}, and {Jones}}]{Salathe}
{Salathe}, M., {Kazandjieva}, M., {Lee}, J.~W., {Levis}, P., {Feldman}, M.~W.,
  {Jones}, J.~H., Dec. 2010. A high-resolution human contact network for
  infectious disease transmission. Proceedings of the National Academy of
  Science 1072, 22020--22025.

\bibitem[{Shalizi and Thomas(2010)}]{Shalizi:2010}
Shalizi, C.~R., Thomas, A.~C., 2010. Homophily and contagion are generically
  confounded in observational social network studies. Sociological Methods
  Research 40~(2), 27.
\newline\urlprefix\url{http://arxiv.org/abs/1004.4704}

\bibitem[{Shrum et~al.(1988)Shrum, Cheek, and Hunter}]{Shrum}
Shrum, W., Cheek, N.~H., Hunter, S.~M., 1988. Friendship in school: Gender and
  racial homophily. Sociology of Education 61~(4), 227--239.
\newline\urlprefix\url{http://www.jstor.org/stable/2112441}

\bibitem[{{SocioPatterns}(2012)}]{SocioPatterns}
{SocioPatterns}, 2012. \url{http://www.sociopatterns.org/}, accessed on June 12
  2012.

\bibitem[{Steglich et~al.(2010)Steglich, Snijders, and Pearson}]{Steglich}
Steglich, C., Snijders, T. A.~B., Pearson, M., 2010. Dynamic networks and
  behavior: separating selection from influence. Sociological methodology,
  329--393.

\bibitem[{Stehl\'e et~al.(2011)Stehl\'e, Voirin, Barrat, Cattuto, Isella,
  Pinton, Quaggiotto, Van~den Broeck, R\'egis, Lina, and Vanhems}]{PlosOne}
Stehl\'e, J., Voirin, N., Barrat, A., Cattuto, C., Isella, L., Pinton, J.-F.,
  Quaggiotto, M., Van~den Broeck, W., R\'egis, C., Lina, B., Vanhems, P., 08
  2011. High-resolution measurements of face-to-face contact patterns in a
  primary school. PLoS ONE 6~(8), e23176.

\bibitem[{Szell et~al.(2010)Szell, Lambiotte, and Thurner}]{szell}
Szell, M., Lambiotte, R., Thurner, S., 2010. Multirelational organization of
  large-scale social networks in an online world. PNAS 107, 13636.
\newline\urlprefix\url{doi:10.1073/pnas.1004008107}

\bibitem[{Vigil(2007)}]{Vigil}
Vigil, J., 2007. Asymmetries in the friendship preferences and social styles of
  men and women. Human Nature 18, 143--161, 10.1007/s12110-007-9003-3.

\end{thebibliography}
\end{document}